\documentclass[reprint,amsmath,amssymb,aps,prl,superscriptaddress]{revtex4-1}

\usepackage{graphicx}
\usepackage{dcolumn}
\usepackage{bm}
\usepackage{hyperref}

\usepackage{xcolor}

\usepackage{subcaption}
\newcommand{\labelphantom}[1]{%
{\phantomsubcaption%
\label{#1}}%
}%

\begin{document}

\title[Vertical Control of SPARC Discharges]{Implications of Vertical Stability Control on the SPARC Tokamak} 

\author{A. O. Nelson}
\email[Corresponding author: ]{a.o.nelson@columbia.edu}
\affiliation{Columbia University, New York City, NY, USA}

\author{D. T. Garnier} 
\affiliation{Massachusetts Institute of Technology, Cambridge, MA, USA}

\author{D. J. Battaglia} 
\affiliation{Commonwealth Fusion Systems, Devens, MA, USA}

\author{C. Paz-Soldan}
\affiliation{Columbia University, New York City, NY, USA}

\author{I. Stewart}
\affiliation{Columbia University, New York City, NY, USA}

\author{M. Reinke} 
\affiliation{Commonwealth Fusion Systems, Devens, MA, USA}

\author{A. J. Creely} 
\affiliation{Commonwealth Fusion Systems, Devens, MA, USA}

\author{J. Wai}
\affiliation{Commonwealth Fusion Systems, Devens, MA, USA}

\vspace{10pt}
\date{\today}

\begin{abstract}
To achieve its performance goals, SPARC plans to operate in equilibrium configurations with a strong elongation of $\kappa_\mathrm{areal}\sim1.75$, destabilizing the $n=0$ vertical instability. However, SPARC also features a relatively thick conducting wall that is designed to withstand disruption forces, leading to lower vertical instability growth rates than usually encountered. In this work, we use the \texttt{TokSyS} framework to survey families of accessible shapes near the SPARC baseline configuration, finding maximum growth rates in the range of $\gamma\lesssim100\,$s$^{-1}$. The addition of steel vertical stability plates has only a modest ($\sim25\%$) effect on reducing the vertical growth rate and almost no effect on the plasma controllability when the full vertical stability system is taken into account, providing flexibility in the plate conductivity in the SPARC design. Analysis of the maximum controllable displacement on SPARC is used to inform the power supply voltage and current limit requirements needed to control an initial vertical displacement of $5\%$ of the minor radius. From the expected spectra of plasma disturbances and diagnostic noise, requirements for filter latency and vertical stability coil heating tolerances are also obtained. Small modifications to the outboard limiter location are suggested to allow for an unmitigated vertical disturbance as large as $5\%$ of the minor radius without allowing the plasma to become limited. Further, investigations with the 3D \texttt{COMSOL} code reveal that strategic inclusion of insulating structures within the VSC supports are needed to maintain sufficient magnetic response. The workflows presented here help to establish a model for the integrated predictive design for future devices by coupling engineering decisions with physics needs. 
\end{abstract}

\vspace{2pc}

\maketitle


\section{Introduction}

In order to maximize fusion performance, tokamak plasmas are typically run with high elongation $\kappa_\mathrm{areal}$, which increases both the plasma volume and the energy confinement time ($\tau_\mathrm{e}\sim\kappa_\mathrm{areal}^{0.7}$) \cite{iter_physics_expert_group_on_confinement_and_transport_chapter_1999}. Here we use the areal elongation $\kappa_\mathrm{areal}\equiv S_\mathrm{0}/(\pi a^2)$, where $S_\mathrm{0}$ is the plasma cross-sectional area and $a$ is the minor radius to be consistent with scaling work done by the ITER Group \cite{ITER1999} rather than the nominal elongation $\kappa$, defined as the cross-sectional height over the width. Unfortunately, while enhancing plasma performance, increasing $\kappa_\mathrm{areal}$ can also act to destabilize magneto-hydrodynamic (MHD) modes like the $n=0$ resistive wall mode, which can trigger vertical displacement events (VDEs) and ultimately result in the violent termination of the plasma discharge. To avoid VDEs while still operating at the maximum achievable elongation, many tokamaks employ passive stabilization techniques \cite{leuer_passive_1989, lee_numerical_1999, ferrara_plasma_2008, buxton_design_2018} and/or active real-time control feedback loops \cite{humphreys_diii-d_2008, humphreys_experimental_2009, xue_hot_2019, pesamosca_improved_2022} that work to prevent the uncontrolled growth of the vertical instability.

In support of experimental endeavours to establish reliable vertical control of tokamak plasmas, several modeling suites have been developed that can accurately capture the interplay between the plasma state, the vessel geometry and the coil power supplies \cite{humphreys_development_2007, welander_closed-loop_2019, hansen_tokamaker_2023, artola_3d_2021, fitzpatrick_simple_2009}. Since the physics governing VDEs is relatively well understood \cite{lazarus_control_1990, portone_stability_2005, nelson_vertical_2023, Sweeney2020, yolbarsop_analytic_2022}, it is possible to use these codes to predict the expected stability margins for future devices with fairly good accuracy. This is especially relevant for the devices like SPARC, which require rapid and high-fidelity assessment of numerous and often competing physics and engineering challenges in order to facilitate design and construction of the device on an accelerated timescale \cite{Creely2020, Sweeney2020, stewart_optimization_2023, battey_design_2023}. For example, related modeling work on SPARC has previously led to modifications of the runaway electron mitigation coil \cite{battey_design_2023} and has allowed for optimization of the installation locations for magnetic sensors that will inform equilibrium reconstructions and vertical control measurements \cite{stewart_optimization_2023}.

In this work, we utilize the \texttt{TokSys} code base \cite{humphreys_development_2007, welander_closed-loop_2019} to assess the implications of vertical stability control for the SPARC tokamak. In section~\ref{sec:passive}, we develop and apply a realistic model of the SPARC device using the \texttt{GSPert} code collection to assess the passive stability of expected plasmas in SPARC, ultimately informing design decisions regarding the shape of the limiting surface. Then, in section~\ref{sec:active}, we introduce models of the vertical stability control system to understand limitations on the controllable vertical displacements that may occur naturally during operation. These results indicate that a robust vertical control system capable of withstanding seed disturbances from a variety of expected sources can be established on SPARC. Finally, we allow several aspects of this control system to vary to demonstrate the impact that certain free parameters will have on SPARC operation in section~\ref{sec:cntrl}, providing input into the optimization of power supply voltages, currents and latencies. A short summary of conclusions and lessons learned for other devices are presented in section~\ref{sec:conc}.


\section{Passive Stability of SPARC Discharges}
\label{sec:passive}

Calculation of the natural instability growth rates of tokamak equilibria is generally a tractable problem, as long as an accurate description of the machine in question is available \cite{humphreys_development_2007, hansen_tokamaker_2023}. In this work we use a collection of calculations known as \texttt{GSPert}, which are included in the \texttt{TokSys} code suite, to compute the open-loop growth rates ($\gamma$) of the $n=0$ vertical instability for SPARC discharges \cite{humphreys_development_2007, welander_closed-loop_2019}. A 2D model of the machine geometry, as employed in the \texttt{GSPert} model, is shown in figure~\ref{fig:geo}. Within this model, the growth rate is computed including the effect of induced currents in passive conducting structures by employing a linearized Grad-Shafranov solution with a fixed current distribution and a rigid boundary to determine the most unstable eigenmodes for each configuration, following similar studies on other machines \cite{nelson_vertical_2023, humphreys_experimental_2009, liu_controllability_2014, qiu_comparison_2017}. These calculations can be extended to track the time-dependent motion of the rigid boundary in response to time-dependent currents and/or voltages in the active coils, as will be discussed in section~\ref{sec:active}.

\subsection{Machine Description}
\label{subsec:machine}

\begin{figure}
	\includegraphics[width=1\linewidth]{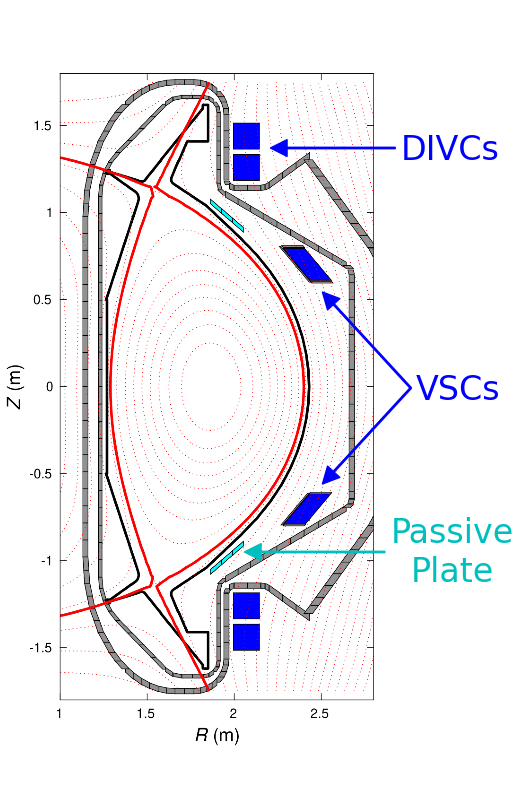}
	\caption{The nominal SPARC scenario discussed in this work, from \cite{creely_sparc_2023}. Vertical stability coils (blue -- VSCs) and candidate passive plates (cyan) are both included inside of the conducting SPARC shell (grey) for increased response, whereas divertor control coils (blue -- DIVCs) and other shaping control coils (not shown) are located outside of the vacuum vessel. The nominal equilibria discussed in this work is shown in red. }
	\label{fig:geo}
\end{figure}

To provide the required time response to minimize large deviations in the vertical position, the vertical stability coils (VSCs) on SPARC are composed of copper coils and located inside the conducting vacuum vessel, as shown in figure~\ref{fig:geo}. The upper and lower coil sets are wired in anti-series in order to produce a quadrupole field that applies a vertical force on the plasma. The plasma position will be inferred via magnetic measurements, with possible enhancements from optical measurements \cite{stewart_optimization_2023}. The desired VSC current delivered by the power supply will be calculated within a stand-alone plasma control system by comparing the inferred position and velocity of the plasma vertical motion in relation to a target position, typically defined as $Z_\mathrm{p}=0$, where $Z_\mathrm{p}$ is the plasma magnetic axis.

In addition to the VSCs, the central solenoid system, poloidal field system and divertor coil (DIVCs) sets on SPARC can also influence the vertical position of the plasma since each set has an upper and lower coil that can be controlled independently. However, these systems are outside of the thick, double-walled vacuum vessel and are thus are only capable of influencing the magnetic field in the plasma volume at a rate that is at least an order of magnitude slower than the VSCs. As such, only the impact of the VSCs is assessed in this work. In practice, the vertical controller will integrate fast and slow commands over all of the coils in order to maintain the average VSC current near zero over long ($\sim100\,$ms) timescales. Long-time-scale controller errors, such as integrator drift in the magnetic measurements and sensor motion due to thermal expansion of structures, can lead to errors in the positioning of the plasma, and should be mitigated during SPARC operation.

\begin{figure*}
    \includegraphics[width=0.9\linewidth]{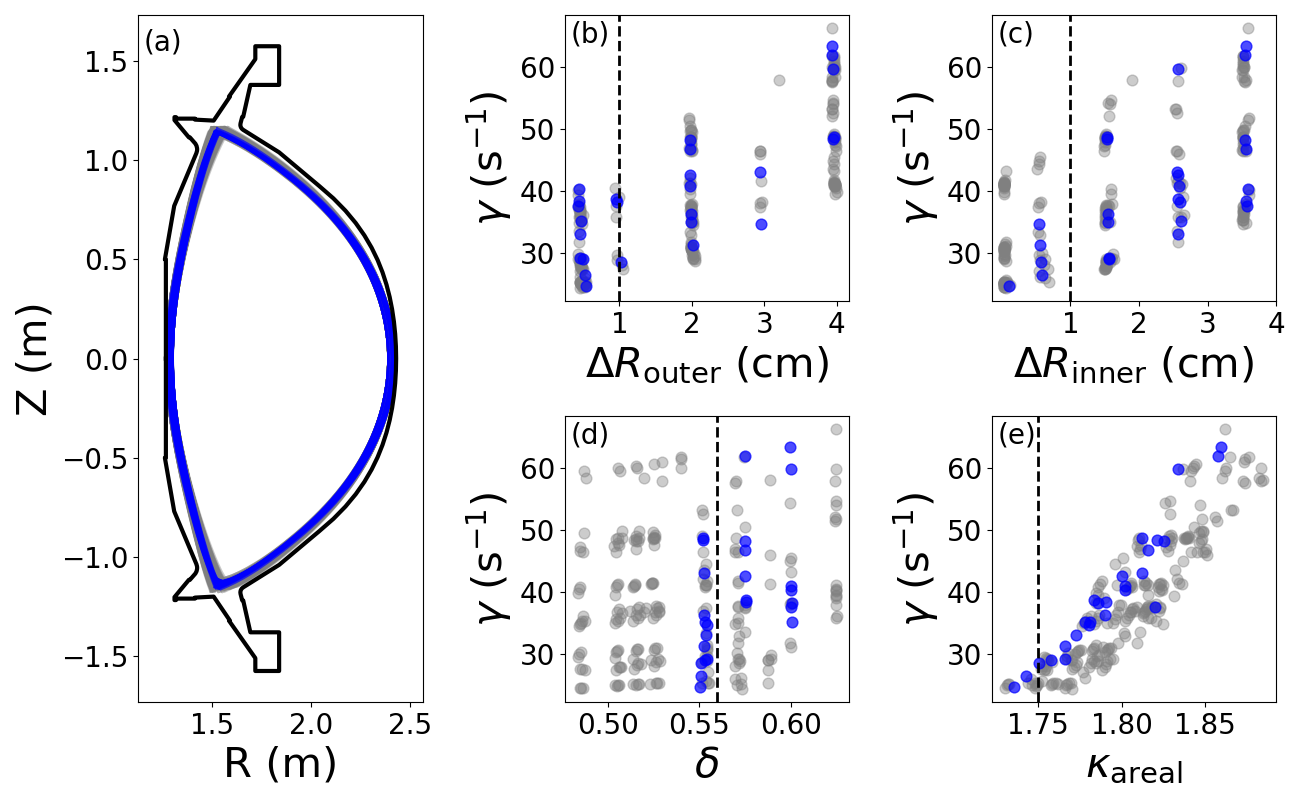}
    \labelphantom{fig:paramspace-a}
    \labelphantom{fig:paramspace-b}
    \labelphantom{fig:paramspace-c}
    \labelphantom{fig:paramspace-d}
    \labelphantom{fig:paramspace-e}
    \caption{(a) A depiction of one of the families of equilibria used in this work. The growth rate $\gamma$ is shown for each equilibria as a function of in (b) outer gap, (c) inner gap, (d) triangularity and (e) elongation. Points colored in blue feature divertor-compatible strike lines that land on the high-heat-flux plasma-facing components. In (b-e), dashed vertical lines represent the nominal operating point from figure~\ref{fig:geo}. 
    }
    \label{fig:paramspace}
\end{figure*}

The remainder of the 2D model shown in figure~\ref{fig:geo} includes a description of both vacuum vessel shells using the material resistance of the nitronic steel. The model description also includes a shell around the VSC to capture the effect of induced current in the VSC supporting structures and VSC jacket. The effective resistance of the VSC supporting structure elements was tuned to provide good agreement with a higher-fidelity 3D model produced in \texttt{COMSOL} \cite{comsol_inc_comsol_2013}, as discussed in section~\ref{subsec:3D}. Continuous passive plates made of stainless steel can also be included inside the vacuum region near the DIVC coils. The vertical stabilizing eigenmode of this 2D vessel description is $\tau_\mathrm{wall}\sim47\,$ms with passive plates and $\tau_\mathrm{wall}\sim45\,$ms without the plates, indicating that the thick conducting walls of SPARC already account for most of the passive stabilization achieved by the device. Because of this, the final SPARC design does not include include these plates as originally conceived, as explained below. These times match wall times calculated with a 3D model completed with the Psi-Tet code \cite{hansen_numerical_2015, battey_design_2023}. Further discussions on the effect of passive plates is included in section~\ref{subsec:plates}.


\subsection{Analysis of Equilibria Database}
\label{subsec:eqdb}

In order to best assess the implications of vertical stability control in a predictive manner, several ``families'' of equilibria are generated for analysis that feature variation in key parameters around various nominal operating points. These include sets of discharges surrounding the nominal SPARC H-mode scenario know as the ``primary reference discharge" \cite{creely_sparc_2023} (depicted in figure~\ref{fig:geo}) or surrounding various transient conditions expected during the ramp-up and ramp-down phases of SPARC pulses. Each equilibrium included in this analysis can be generated with the planned SPARC coil locations and current limits. Figure~\ref{fig:paramspace} provides a summary of a database constructed around the nominal $Q>1$ operating point, highlighting the impact of various shaping parameters on the nominal VDE growth rate $\gamma$. A subgroup of this equilibria family is marked in blue, indicating double-null configurations which place the strike points on the portion of the SPARC divertor designed to tolerate high heat fluxes \cite{Creely2020, Kuang2020}. This constraint will be enforced during operation, constraining the shaping of SPARC scenarios to a smaller range than considered in this study. Across the range of shapes shown in figure~\ref{fig:paramspace}, the growth rates do not exceed $\gamma\sim70\,$s$^{-1}$, which, given the $\tau_\mathrm{wall}$ values reported above, is typically well within the controllable range seen existing devices  \cite{ferrara_plasma_2008, humphreys_diii-d_2008, humphreys_experimental_2009}. 

From these database analyses, several important trends for stability control on SPARC are already demonstrated. Figures~\ref{fig:paramspace-b} and \ref{fig:paramspace-c} show the effect of variations in the outer gap and inner gap on $\gamma$, respectively. In both cases, increased separation between the midplane wall and the separatrix increases the growth rate, as is expected due to the less effective wall coupling and general increase in elongation. During SPARC operation, these values will likely be constrained by coupling to the RF power systems \cite{Lin2020} and the desire to maximize the plasma volume, and will thus not be the primary leveraging knobs for stability control. The triangularity (figure~\ref{fig:paramspace-d}) has very little effect on $\gamma$, separating the vertical stability issue from that of strike point sweeping during high power operation \cite{Kuang2020}. Notably, the boundary elongation ($\kappa_\mathrm{areal}$) has the largest direct impact on $\gamma$ for a given configuration and will thus be the primary control level used to mitigate any emergent stability concerns. This in part informs the selection of $\kappa_\mathrm{areal}=1.75$ as a target for nominal operation. Variations in internal parameters such as the normalized beta ($\beta_\mathrm{N}$) and internal inductance ($l_\mathrm{i}$), which are well known to impact the maximum allowable elongation \cite{gates_progress_2006, jin-ping_stability_2009}, are also included in these databases. However, since their impact on $\gamma$ is generally dominated by the parameters describing wall separations shown in figure~\ref{fig:paramspace} and their values will ultimately be set by the core SPARC mission rather than directly as control variables, we consider modification of $\beta_\mathrm{N}$ and $l_\mathrm{i}$ here only as natural variation around the baseline scenario. Overall, these results point to the importance of accurate and stable real-time control of the inner gap position and X-point position for divertor compatibility and vertical stability control. 

Note that separate equilibria databases were created with larger inner and outer gaps and reduced X-point vertical $|Z|$ positions. These test double-null shapes that stay diverted but pull away from the wall in order to qualify stability over a series of equilibria that may potentially be employed during plasma ramp-down. Due to reduced coupling with the wall, these equilibria generally yield larger growth rates, above $\gamma\sim100\,$s$^{-1}$ in some cases. However, further optimization of the ramp-down strategy in SPARC is still ongoing, so interpretations of vertical stability control during this transient evolution is not included in detail here. 

\subsection{Implications for Limiter Surface Shape}
\label{sec:pfcs}

Since these equilibria families were computed early in the design process for SPARC, it is possible to utilize them to inform tangible design decisions. Due to its reliance on ion cyclotron resonant frequency (ICRF) heating, the nominal SPARC operating point will likely need to sustain a relatively small outer gap on the order of $\sim2\,$cm to enable efficient edge coupling \cite{Lin2020}. This presents a challenge for control, as even small radial perturbations (and small vertical perturbations in the case of a conformal wall) could be sufficient to induce a limited plasma state during full-power operation. Due to the large amount of stored energy present in higher performance SPARC discharges, limited operation is not foreseen as a material-compatible operation space and must therefore be avoided.

\begin{figure}
	\includegraphics[width=0.9\linewidth]{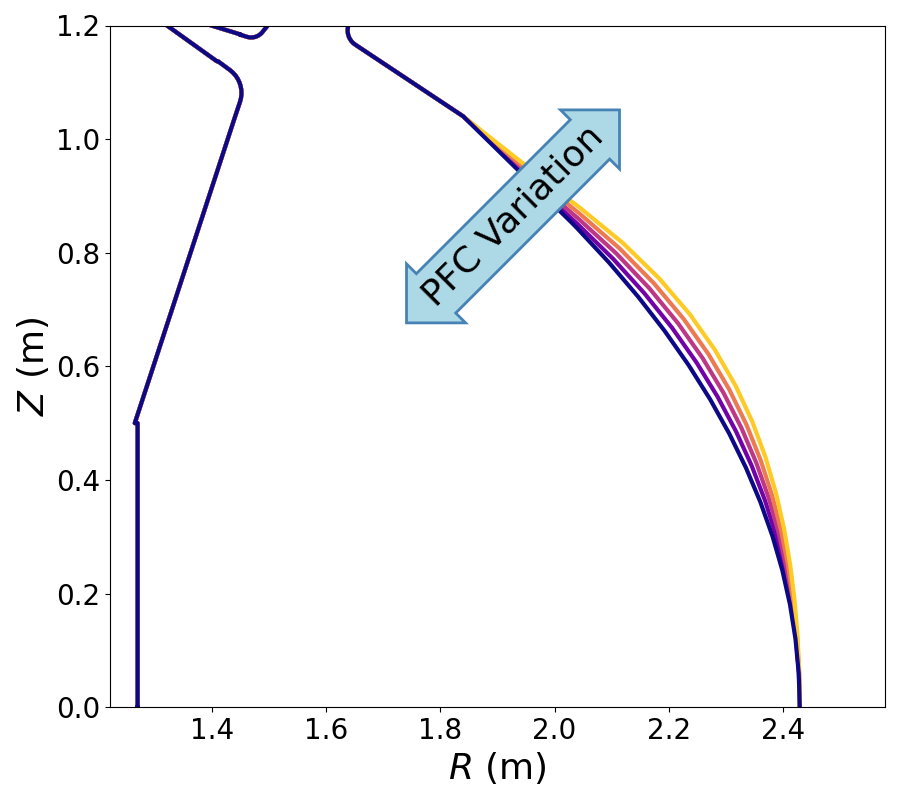}
	\caption{Potential options for the SPARC PFC contour. Colors correspond to the analysis presented in figure~\ref{fig:zcoll_pfc}.}
	\label{fig:pfcs}
\end{figure}

To mitigate this risk, variations of the plasma-facing component (PFC) contour were assessed based on their compatibility with vertical control. The outboard limiter on SPARC consists of 18 independent poloidal rail limiters, each with a few degrees of toroidal width, which transitions to a toroidally continuous upper and lower off-midplane limiters. The shape of the off midplane limiters are conical surfaces, constrained generally by the divertor and vacuum vessel shape. While the low-field-side limiter point was constrained as part of the early radial build specification of the device, the poloidal rail limiter shape between $Z=0$ and the off-midplane conical surface is free to be specified by physics needs. A polynomial parameterization was chosen to allow for easy calculation of local slope along the contour, which could then be transferred to the design of steel structural components which approximate the curve with flat surfaces. The up-down symmetry of SPARC forced this polynomial to be even and a continuity requirement that the slope of the polynomial match the slope of the off-midplane conical surface left a single variable for an $n=4$ polynomial.  Limiter shapes included in this study were scanned by adjusting, within a few cm's of $|Z|$, the poloidal rail and conical limiter surface transition point, resulting in the variations depicted in figure~\ref{fig:pfcs}. An $n=6$ polynomial was also considered, but there was neither a large need nor sufficient physics to constrain a 2D parameter search.

\begin{figure}
	\includegraphics[width=1\linewidth]{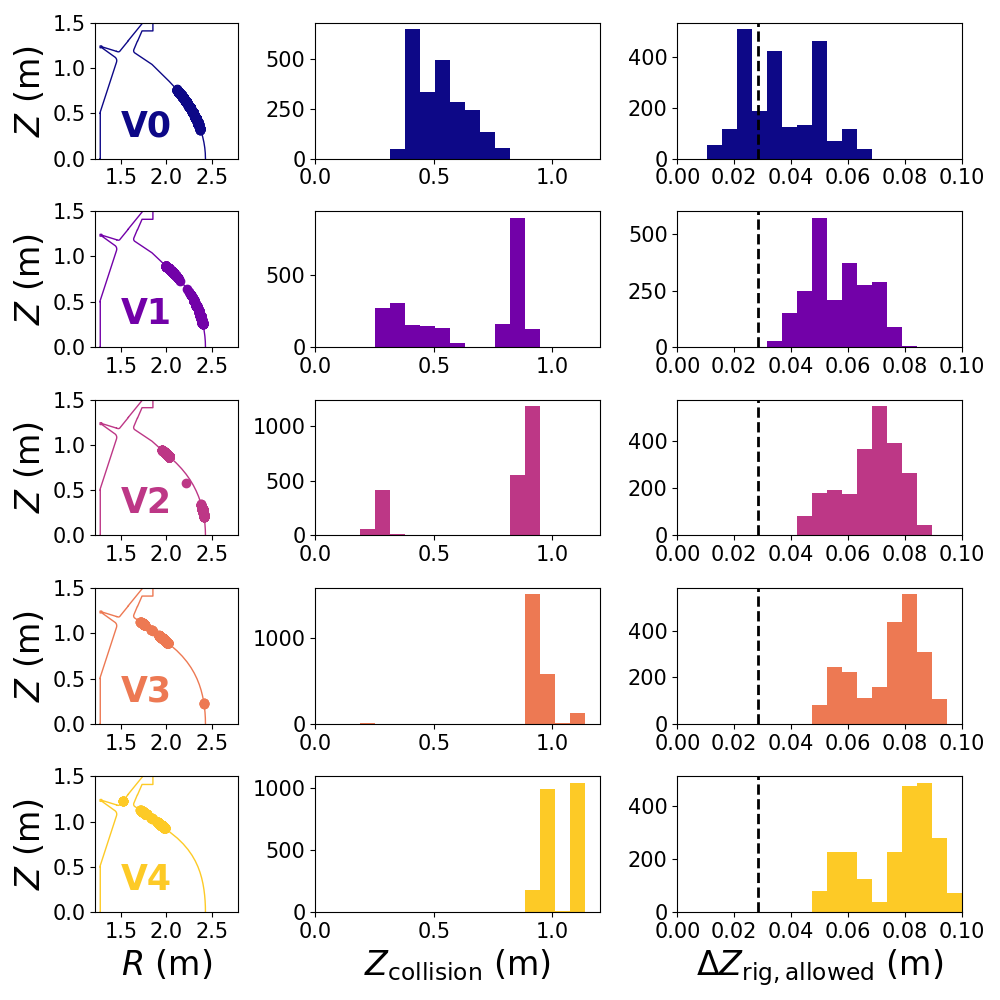}
	\caption{Impact of PFC contour shape on collision rate due to vertical displacements, for the five potential shapes considered in figure~\ref{fig:pfcs}. Left column: impact location of each equilibria along the PFC contour. Middle column: height of each impact location. Right column: Maximum vertical displacement ($\Delta Z_\mathrm{rig,allowed}$) achieved before each impact. The value of $\Delta Z_\mathrm{rig,allowed}$ corresponding to $\Delta Z_\mathrm{max}/a = 5\%$ is marked with a vertical dashed line.}
	\label{fig:zcoll_pfc}
\end{figure}

For each of these PFC variations, the nominal operating equilibria shown in figure~\ref{fig:paramspace} were rigidly shifted in the plasma volume until intersection between the last closed flux surface and the limiter wall was achieved. During actual operation, the plasma shape will deform in response to image currents as it approaches the conducting wall, so this rigid vertical shift is representative of a ``worst-case" operational safety margin for each scenario. The total vertical distance that each equilibria can travel before colliding with the wall ($\Delta Z_\mathrm{rig,allowed}$) is shown for each PFC option in figure~\ref{fig:zcoll_pfc}. Most importantly, histograms of $\Delta Z_\mathrm{rig,allowed}$ for the considered equilibria are provided in the right-most column of figure~\ref{fig:zcoll_pfc} for a variety of different limiter surfaces, which are colored based on the depictions in figure~\ref{fig:pfcs}. It is initially evident that, as the curvature of the limiting surface is increased from version ``V0'' (purple) to version ``V4" (orange), $\Delta Z_\mathrm{rig,allowed}$ increases for a large portion of the database. Further, the equilibria are more likely to intersect the limiting surface near the divertor region at higher PFC curvature, which potentially offers an opportunity to add high heat flux materials in the region near the outboard divertor slot in anticipation of more frequent plasma contact. However, it is not sufficient to simply adopt the PFC contour shape that produces the largest contained volume -- this both reduces the wall coupling responsible for passive stabilization of the SPARC equilibria and reduces the available space directly outside of the PFC structure that is needed for additional engineering components. A close assessment of figure~\ref{fig:zcoll_pfc}, reveals that only marginal gains in $\Delta Z_\mathrm{rig,allowed}$ are achieved past an option with more moderate curvature between the ``V1" and ``V2" options, indicating an optimal PFC contour arrangement. Further, we note that the smallest values of $\Delta Z_\mathrm{rig,allowed}$ achieved with this configuration are still larger than $5\%$ of the minor radius (marked with a vertical dashed line), which is found to be the critical margin for safe operation below. 


\section{Active Stabilization of Rigid Displacements}
\label{sec:active}

Passive analysis of the nominal SPARC operating scenario optimistically suggests good controllability of the vertical instability. However, off-normal and dynamic events can place more rigorous demands on the VSC system. Assessment of these events requires time-dependent analysis, which can be developed by extending the \texttt{TokSys} models with a simple vertical controller, as is presented below.

\subsection{Vertical Controller}
\label{subsec:PID}

For the purposes of this scoping work it is sufficient to employ the simplest of fast vertical controllers, which uses proportional-derivative (PD) feedback on the difference and rate-of-change of the difference between the inferred vertical position of the plasma and the target vertical position. We do not yet include a diagnostic model in this work (though the impact of measurement error is discussed below), so the inferred vertical position is set to be equal to the true equilibrium position at all times. For the PD controller, we choose reasonable values for the proportional feedback coefficients ($K_\mathrm{prop} = -33\,$kV/m) and derivative ($K_\mathrm{deriv} = -0.156\,$kV/(m/s)) in order to critically damp the $Z$-position for the modeled equilibria. More rigorous development and tuning of the position controller, including the deployment of more advanced algorithms, will be pursued in the future.

During operation of SPARC, magnetic measurements fed to the fast vertical controllers will be filtered in order to reduce the influence of noise on the vertical position estimation. This filtering introduces a latency into the system, which is assumed to be characterized by the time constant $\tau_\mathrm{filter}=0.2\,$ms. The sensitivity of these results to this latency choice is discussed in more detail in section~\ref{sec:cntrl}. Additionally, we include a $0.3\,$ms delay in applying the voltage to capture other sources of lag, which accounts for communication and processing delays in the control loop. This effectively increases the $\tau_\mathrm{filter}$ parameter -- the inclusion of additional time constants into the model all have the same impact on the vertical controller. As such, all potential delay times can be captured through a single delay constant. Finally, we note that power supply voltage droop, which is a drop in output voltage from the power supply as it drives a load at large currents, is not included in this model. Instead, operation away from the power supply voltage limit ($1.2\,$kV) is emphasized as a potential strategy to improve the chances that the droop will not impact the voltage applied to the VSC.

\subsection{Vertical Perturbation Size}
\label{subsec:dZ0}

One of the largest uncertainties in this sort of predictive modeling is in quantifying the size of the initial seed for vertical displacement events, which can come from a variety of sources. Since the height of this initial perturbation can significantly impact both the ability to control the plasma and the voltages needed to do so, each potential source was considered separately to establish a range of expected vertical seeds on SPARC. 

\textit{Noise} -- Measurement noise, signal interference or non-axisymmetric fields or currents can be incorrectly interpreted as vertical motion in real time, leading to an errant response from the controller.  Additionally, small imbalances in the current (including transients) from upper and lower coil pairs can also introduce vertical perturbations. Previous studies have quantified the typical deviation in vertical position signal measurement due to noise and disturbance on a range of small to mid-sized tokamaks, including Alcator C-Mod at high field, finding that the typical size of this effect is $0.2\% \lesssim \Delta Z / a \lesssim 1.5\%$ \cite{humphreys_experimental_2009}. While the impact of noise can potentially be ameliorated via a careful systematic selection of sources or via low pass filtering, this could introduce increased latency into the system leading to further complications in the employed control schemes \cite{humphreys_experimental_2009}. For consideration in this work, we assume a $3\sigma$ distribution for potential disturbances from noise, indicating that the largest expected perturbations from noise on SPARC should be on the order of $0.3 - 2.7\,$cm ($\Delta Z / a=0.5-4.7\%$). 

\textit{Rapid Equilibrium Change} -- The rapid ($\sim\,$ms) redistribution of current in the plasma or the induced current in passive structures can also perturb the vertical position of the plasma. This sort of current instability can occur from MHD instabilities \cite{Sweeney2020}, edge-localized modes (ELMs) \cite{nelson_time-dependent_2021, ali-arshad_plasma_1996, villone_position_2005}, sawteeth \cite{chapman_controlling_2011}, L-H and H-L transitions \cite{nelson_experimental_2020}, and/or the rapid termination of the RF or nuclear heating \cite{walker_emerging_2006}, among others. To quantify this effect, $10\%$ drops in $\beta_\mathrm{p}$ and $l_\mathrm{i}$ and $1\%$ drops in $I_\mathrm{p}$ were modeled in both double and single null SPARC equilibria, leading to sub-mm changes in the vertical position. This small vertical effect is a result of the significant inductance and poloidal uniformity of the vacuum vessel, which minimizes the size and up/down asymmetry of the induced currents. We note that radial perturbations observed during these events were much larger (on the order of $1 - 2\,$cm), suggesting that rapid changes in the core plasma may still pose a challenge for maintaining small inner and outer wall gaps.

\textit{ELMs} -- Of course, ELMs themselves can also directly produce significant vertical perturbations. Most experimental observations of this effect show a rapid motion toward the dominant X-point followed a slow recoil back to the normal operating position, consistent with the transfer of current to the scrape-off-layer (SOL) via peeling modes being concentrated near the primary X-point \cite{ali-arshad_plasma_1996}. This has led to maximum excursions on the order of $\Delta Z / a \lesssim 5-10\%$ on JET \cite{ali-arshad_plasma_1996} and $\Delta Z / a \sim 3\%$ on DIII-D \cite{humphreys_experimental_2009}. Since SPARC is projected to eventually operate near peeling stability limits in H-mode, it is reasonable to assume that ELMs will again produce fast vertical motion. However, the magnitude of the effect may be slightly smaller than on JET or DIII-D due to the close-fitting, highly conductive wall and passive plates, and due to potential variation in the size of the SPARC H-mode pedestal \cite{Hughes2020, nelson_setting_2020}. Furthermore, SPARC may target, or only be able to access, regimes either without ELMs or with small, frequent ELMs, so it is possible that the vertical perturbations from ELMs in SPARC will be proportionally smaller than reported elsewhere.

Based on this information, we assume that most seed perturbations will be less than or equal to $3\%$ of the minor radius, with perturbations as large as $\Delta Z / a \sim 5\%$ ($\pm2.9\,$cm) possible when sources are compounded or if very large ELMs are produced. The perturbations are expected to primarily be driven by sensor noise, power-supply transients, non-axisymmetric fields, and instabilities that redistribute plasma current to the SOL. 

\subsection{Maximum Controllable Displacement}
\label{subsec:dZmax}

\begin{figure}
	\includegraphics[width=1\linewidth]{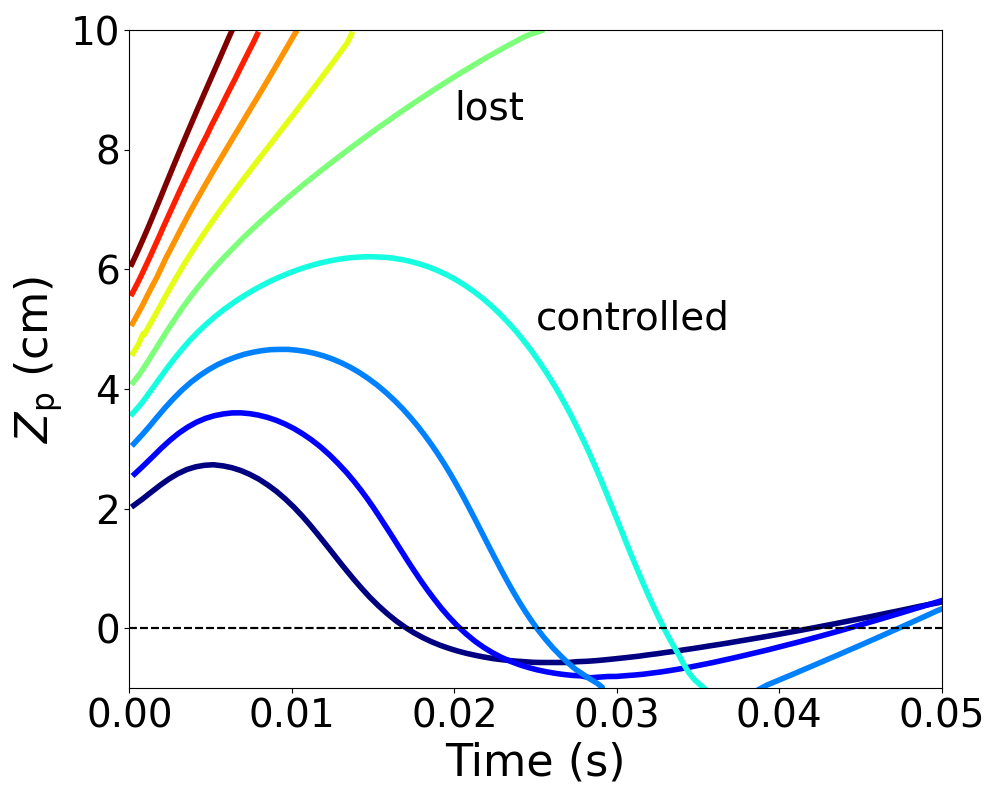}
	\caption{Evolution of the SPARC equilibrium subject to initial vertical perturbations of varying magnitude.}
	\label{fig:z0}
\end{figure}

A common method to characterize the robustness of a vertical control system to these events is to determine the maximum controllable displacement ($\Delta Z_\mathrm{max}$) for a given equilibrium and device description \cite{humphreys_experimental_2009, liu_controllability_2014, qiu_numerical_2019, qiu_comparison_2017}. If $\Delta Z_\mathrm{max}$ is larger than the largest expected vertical excursion, stable operation is likely (though not guaranteed.) Following work by Humphreys, \textit{et al.} \cite{humphreys_experimental_2009}, we calculate $\Delta Z_\mathrm{max}$ using the linear calculations in the \texttt{GSPert} code within \texttt{TokSys} in figure~\ref{fig:z0}. Here initially stable equilibria are rigidly displaced by a distance $Z_\mathrm{0}$ at $t=0$ and the resulting trajectory, which is governed by the vertical growth rate of the initial equilibrium, the power supply (maximum $V$, $I$ and slew rate), controller latency, and the coupled control coils and passive elements that are included in the model. The initial machine conditions are set by changing unstable mode currents by the linear value $x_\mathrm{0}$, defined by
\begin{equation}
    x_\mathrm{0} = \frac{w_\mathrm{z}Z_\mathrm{0}}{w_\mathrm{z}\cdot \frac{dZ}{di_\mathrm{s}}},
\end{equation}
where $w_\mathrm{z}$ is the unstable eigenvector responsible for perturbing the currents in the unstable direction by an amount equivalent to a vertical displacement $Z_\mathrm{0}$ and $\frac{dZ}{di_\mathrm{s}}$ is the proportional change needed in each coil current $i_\mathrm{s}$ to produce a change in the vertical position $Z$. For the particular case shown in figure~\ref{fig:z0}, displaced plasmas with $Z_\mathrm{0}\lesssim4\,$cm are eventually controlled, indicating an approximate measure of $\Delta Z_\mathrm{max}\sim3.5\,$cm for the baseline SPARC scenario.

Normalization of these values to the minor radius provides an effective method with which to compare to existing machines. Multi-machine analysis \cite{humphreys_experimental_2009} of this quantity determined that $\Delta Z_\mathrm{max}/a$ can be interpreted as:
\begin{itemize}
    \item $\Delta Z_\mathrm{max}/a \sim 2\%$ guarantees loss of vertical control;
    \item $\Delta Z_\mathrm{max}/a \sim 4\%$ provides marginal control;
    \item $\Delta Z_\mathrm{max}/a > 5\%$ corresponds to ``safe" operation; and
    \item $\Delta Z_\mathrm{max}/a > 10\%$ corresponds to ``robust" operation.
\end{itemize}
With a minor radius of $a=57\,$cm, figure~\ref{fig:z0} achieves $\Delta Z_\mathrm{max}/a \sim 6-7\%$ on SPARC, implying that safe operation can be expected. Similar metrics ($\Delta Z_\mathrm{max}/a = 5\%$) have been adopted by ITER and demo designs \cite{humphreys_experimental_2009, hahn_advances_2020, villone_coupling_2013}. We note, however, that the maximum excursion of the a controlled trajectory would lead likely to a limited plasma on SPARC, especially if radial motion is included in the disturbance. Therefore, while the maximum controllable displacement criteria is useful as a metric, particularly to compare to other devices, the controllability analysis described in the next section is what drives design decisions.

Several uncertainties remain through this relatively pessimistic approach. Notably, the experimentally achievable $\Delta Z_\mathrm{max}$ could be significantly smaller than the computed $\Delta Z_\mathrm{max}$ due to non-linear effects such as plasma response and equilibrium changes, though this is most often encountered with large perturbation distances, which are not likely in SPARC due to the relatively small minor radius. Non-linear models have been completed for a few representative equilibria, finding results within $10\%$ of the linear model discussed above. It could also be the case that the perturbation spectrum is larger than described here, or that vertical excursions will compound to ratchet the VSC currents away from zero. If prevalent, these effects may eventually require hardware upgrades or scenario down-selection to minimize vertical perturbations, which can be pursued if needed.


\section{Implications for SPARC Design}
\label{sec:cntrl}

Using the framework established above, further details can be added to the vertical stability model in order to both better capture the expected dynamics and to help inform design choices related to the VSC system. 

\subsection{Impact of Controller Hardware Limits on Controllability}

\begin{figure}
	\includegraphics[width=1\linewidth]{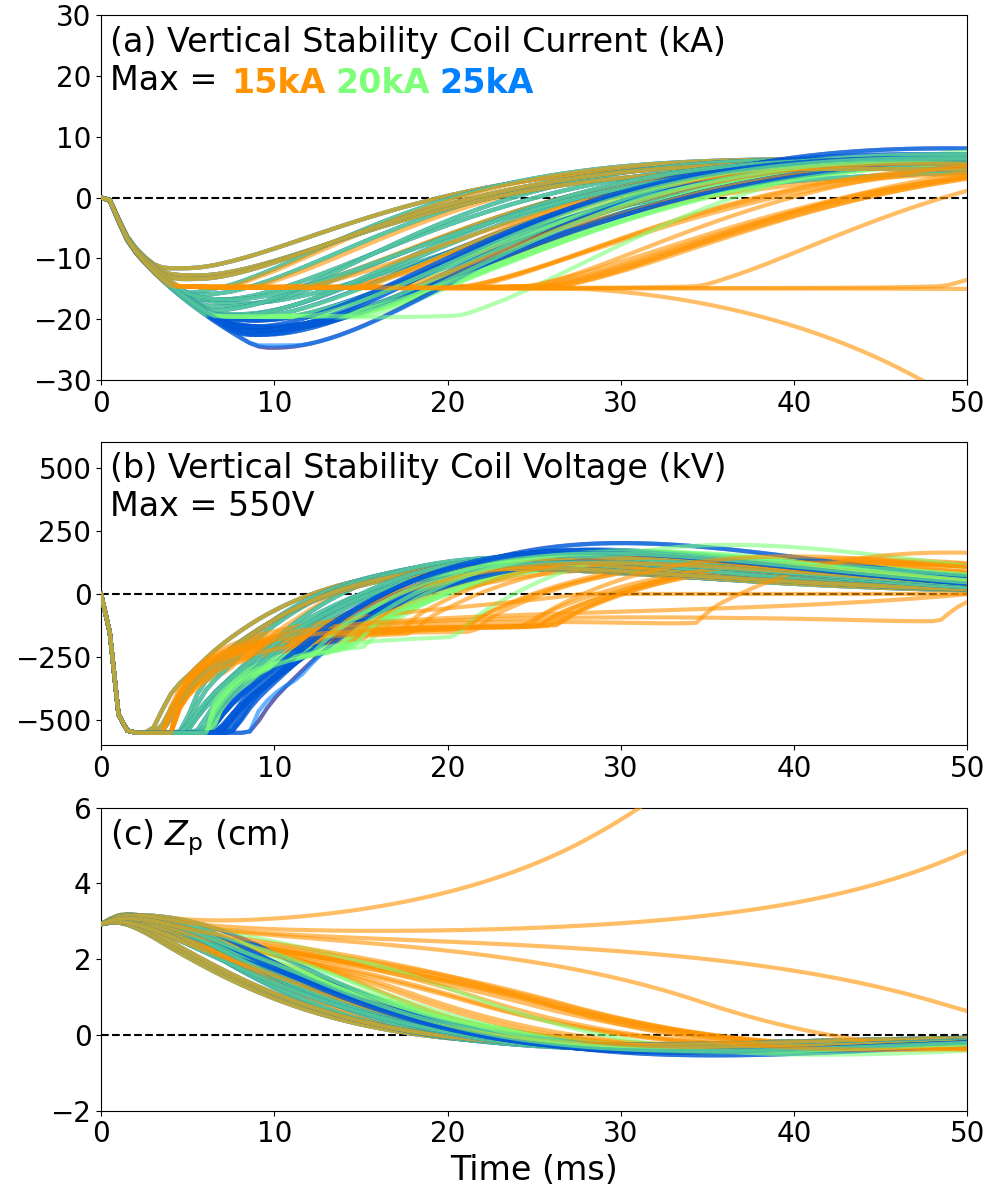}
        \labelphantom{fig:iVSC-a}
        \labelphantom{fig:iVSC-b}
        \labelphantom{fig:iVSC-c}
	\caption{Time traces of various perturbed equilibria subject to vertical stability control under with different VSC current limits: $I_\mathrm{VSC,max}=15\,$kA (orange), $I_\mathrm{VSC,max}=20\,$kA (green), $I_\mathrm{VSC,max}=25\,$kA (blue). The (a) VSC current, (b) VSC voltage, and (c) plasma magnetic axis are shown. }
	\label{fig:iVSC}
\end{figure}

One of the primary leveraging parameters on the success rate of the VSC system is the maximum coil current allowed in the VSCs ($I_\mathrm{VSC,max}$). Using the \texttt{GSPert} model, it is possible to model stability control under several different $I_\mathrm{VSC,max}$ assumptions, as is shown in figure~\ref{fig:iVSC}. Here an entire family of equilibria that satisfy the divertor compatibility criterion is perturbed by $2.85\,$cm (equivalent to $5\%$ of the minor radius) and allowed to evolve subject to the simple vertical stability controller described in section~\ref{subsec:PID}. If the plasma successfully returns to the nominal operating state of $Z_\mathrm{axis}=0$, and if it stays at least $1\,$cm away from the limiter at all times during the return trajectory, the plasma is classified as ``caught." This $1\,$cm offset accounts for the possibility of the plasma equilibria starting off-midplane due to controller error, as described in section~\ref{subsec:dZ0}. In every simulation, the maximum VSC voltage was set to $550\,$V and the signal was filtered with a low-pass filter with a $0.2\,$ms time constant and a $0.3\,$ms delay. The values are chosen to be roughly representative of the final parameters selected for SPARC operation, though improvements in the engineering of these systems would allow for more aggressive control.

Figure~\ref{fig:iVSC-b} shows the evolution of the voltage across the VSC Coil. In all cases, the voltage saturates at the $550\,$V maximum within $2\,$ms, which is imposed by the assumptions of the maximum voltage slew rate of the H-bridge power supplies. The VSC current limits imposed during each trial become evident shortly after saturation of the VSC voltage, with VSC current rising until almost $10\,$ms after the simulation onset when $I_\mathrm{VSC,max}=25\,$kA. Within this time, most of the $Z$-position trajectories start to roll over, with the exception of a few cases at the lowest VSC coil limit. Notably, though lower current limits correspond to a longer recovery time in some cases, there is no different between the final results for the $I_\mathrm{VSC,max}=20\,$kA and $I_\mathrm{VSC,max}=25\,$kA cases, which both catch $100\%$ of the tested equilibria (see figure~\ref{fig:iVSC-c}). However, several scenarios with the largest growth rates are lost when $I_\mathrm{VSC,max}=15\,$kA, suggesting that $I_\mathrm{VSC,max}\gtrsim20\,$kA will be required for safe operation in SPARC, with higher limits allowing for faster recovery times at the given voltage and latency assumptions.

\begin{figure}
	\includegraphics[width=1\linewidth]{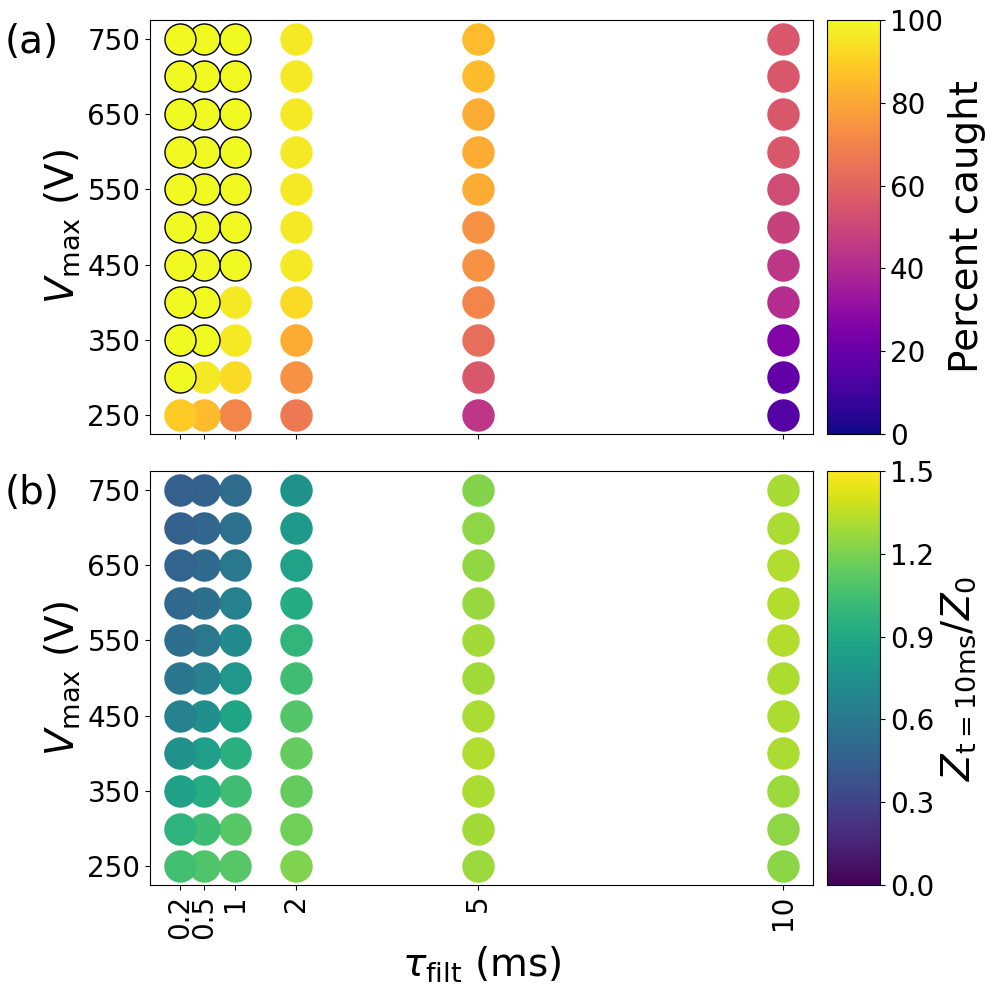}        
        \labelphantom{fig:filt-a}
        \labelphantom{fig:filt-b}
	\caption{(a) The percentage of equilibria that are successfully caught and (b) the average $Z$-position after $10\,$ms of evolution for families of divertor-compatible equilibria subject to vertical stability systems with varying maximum voltage $V$ and time constants $\tau_\mathrm{filt}$. In (a), scenarios in which all equilibria were caught are highlighted with a black circle.}
	\label{fig:filt}
\end{figure}

In a similar manner, the controller latency ($\tau_\mathrm{filt}$) and maximum VSC voltage ($V_\mathrm{max}$) assumptions used above can be assessed via scans over several equilibria families at constant $I_\mathrm{VSC,max}$ limits. We note that other sources of latency can exist in this system, but changing these values has the same general proportional effect as the direct modification of $\tau_\mathrm{filt}$.  The results from this study are summarized in figure~\ref{fig:filt}, which shows the percentage caught and the normalized position after $10\,$ms of evolution ($Z_\mathrm{t=10ms}/Z_\mathrm{0}$, where $Z_\mathrm{0}=2.85\,$cm) as a proxy for the recovery speed. Each circle in figure~\ref{fig:filt} represents about 40 \texttt{GSPert} calculations. In figure~\ref{fig:filt-a}, sets that feature an $100\%$ catch rate are outline in black. This condition was met with a $0.2\,$ms sensor filtering constant if the VSC voltage $V_\mathrm{max}$ is $300\,$V or greater; a $1\,$ms sensor filtering constant is acceptable if $V_\mathrm{max}\gtrsim450\,$V. 

Figure~\ref{fig:filt-b} demonstrates that both the filter latency and the maximum VSC voltage impact the rate at which the plasma position can be restored to the midplane. While there are no direct requirements on this capability from a design perspective, it is generally advisable to restore nominal operation as fast as possible to avoid the potential compounding effects discussed in section~\ref{subsec:dZ0} and to minimize the average VSC current. The voltage has a larger impact on the recovery than the latency, demonstrating some flexibility in the trade-off between maximum voltage and latency around the baseline values of $V_\mathrm{max}=550\,$V and $\tau_\mathrm{filt}=0.2\,$ms. The most significant risk concerning the estimation of required VSC voltage is the possibility of stronger shielding currents in the passive conductors than initially estimated in the model, particularly in regions close to the VSC. This risk is mitigated by predicting the 3D current paths using COMSOL \cite{comsol_inc_comsol_2013}, Psi-Tet \cite{hansen_numerical_2015, battey_design_2023} and CREATE \cite{albanese_linearized_1998} and tuning the conductivity of structures in the VSC model to better match these higher-fidelity models, as discussed in section~\ref{subsec:3D}. 

\subsection{Implications of Passive Plates in SPARC}
\label{subsec:plates}

\begin{figure}
	\includegraphics[width=1\linewidth]{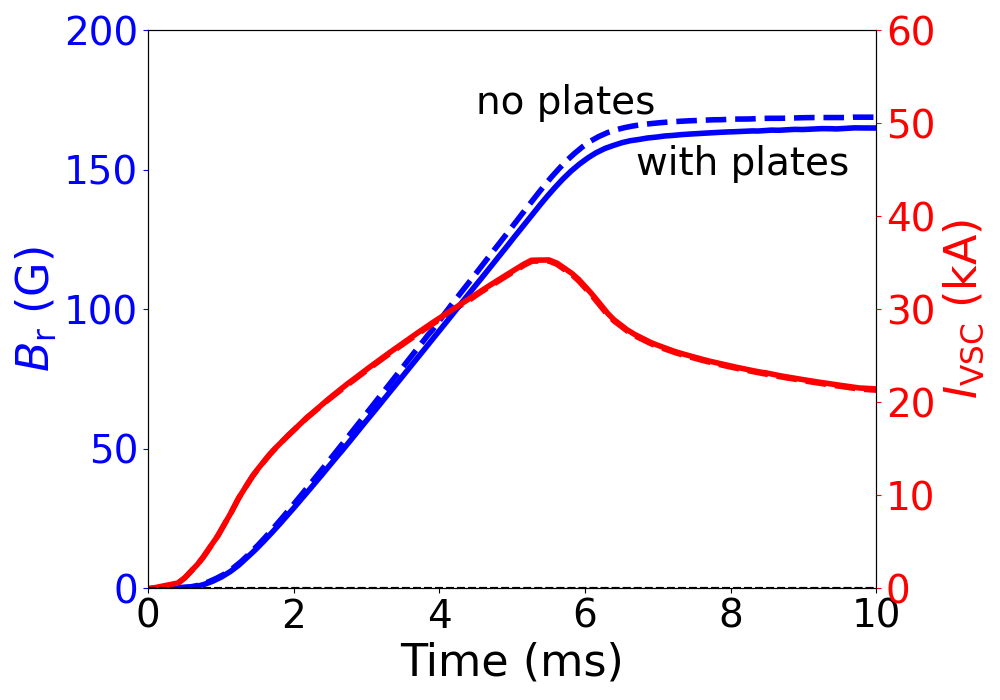}
	\caption{Comparison of radial magnetic field ($B_\mathrm{r}$ -- blue) and VSC current ($I_\mathrm{VSC}$ -- red) for a fixed VSC voltage square wave with (solid lines) and without (dashed lines) the passive plates.}
	\label{fig:plates}
\end{figure}

The impact of stainless steel passive plates on vertical controllability can also be robustly modeled prior to machine operation. The passive plates were removed from the \texttt{TokSys} model by setting the resistance of these elements to a very large number. As seen in figure~\ref{fig:plates}, applying a fixed voltage waveform to the VSC in \texttt{TokSys} results in a similar VSC currents and a slightly larger midplane radial magnetic field without the plates than with the plates. This demonstrates that the induced currents included in the passive plates do lead to some shielding of the VSC field. Further, the nominal growth rate increases slightly from $\gamma=73\,$rad/s to $\gamma=91\,$rad/sec when the continuous stainless-steel plates are removed. 

While the vertical growth rate is larger without the passive plates, the maximum excursion following a $5\%$ perturbation is only slightly larger and, with the same voltage, the plasma $Z$-position is restored slightly faster without plates. This is because removing the passive plates makes the VSC more effective: the coils must no longer fight through the image currents produced by the passive plates, leading to quicker and deeper penetration of the VSC magnetic fields into the plasma volume. Indeed, a repeat of the controllability study presented in section~\ref{sec:active} without plates indicates no dramatic difference in time-dependent behavior with and without the stainless steel plates, though there is a slight reduction in the maximum VSC current without the passive plates. As such, it is concluded that the passive plates in SPARC do not significantly impact the design or performance of the vertical stability system for conductivity ranges less than or equal to stainless steel. The final SPARC design therefore does not include the passive plates as considered in this study. 

Additional studies were performed to assess the impact of other passive plate materials, which can differ significantly from the stainless steel baseline discussed above. Since the resistivity of the plates can be directly modified within this model, it is trivial to quantify variations of stability and time response under different material assumptions. For example, copper stability plates are found to have a large impact on the passive stability of SPARC plasmas due to their higher conductivity. A copper plate, however, would also produced huge forces during a disruption and is therefore difficult to implement on high current, high field systems. As such, inclusion of copper plates is likely not an option for either SPARC or ARC and will not be considered further for the initial designs of these machines. 

\subsection{Impact of 3D Structures on VSC Performance} 
\label{subsec:3D}

As mentioned above, the effective resistances of conducting elements in the 2D \texttt{TokSys} model were tuned in order to provide good agreement with a 3D model developed in \texttt{COMSOL}, which supports high-fidelity modeling of electrical conduction (including induced current) through 3D structures. This allows the 2D model to capture the effect of ports and openings in the vacuum vessel, which would otherwise be missed by the assumption of toroidal symmetry for current-carrying elements. Further, a 2D model can only approximate the overall coupling between the coil and the vessel structure, the effect of which is to reduce the total effective inductance of the coils and the mutual to the plasma. In the absence of vacuum tests shots that can be used for calibration, validation with a 3D model is thus essential for accurate characterization of the vertical stability system. 

\begin{figure}
	\includegraphics[width=1\linewidth]{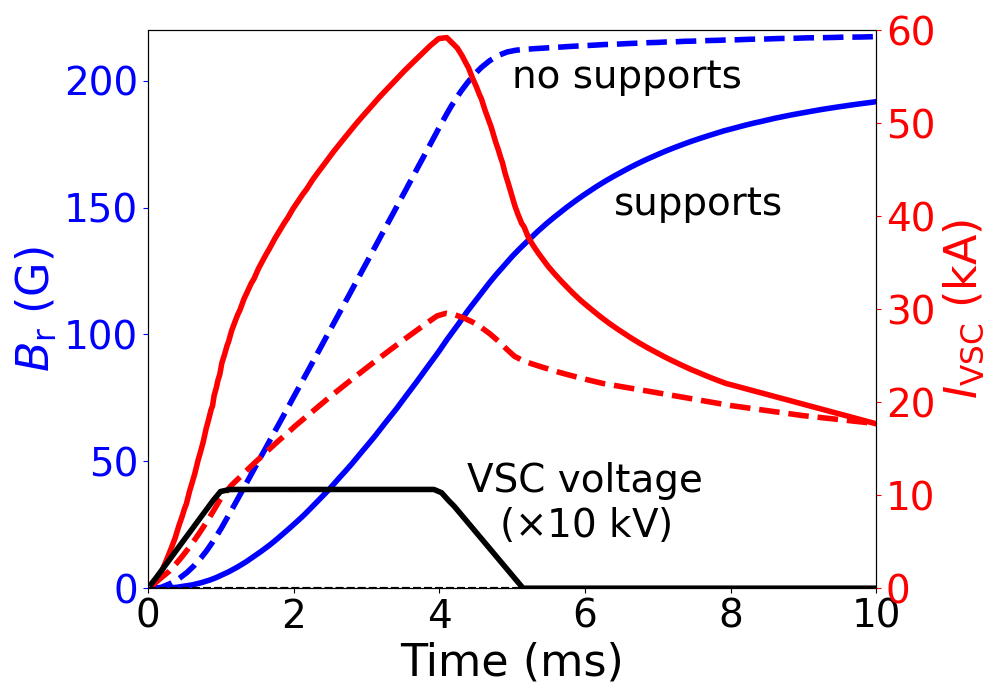}
		\caption{Comparison of radial magnetic field ($B_\mathrm{r}$ -- blue) and VSC current ($I_\mathrm{VSC}$ -- red) for a fixed VSC voltage square wave (black) with (solid lines) and without (dashed lines) the initial design for the supporting structures for VSCs.}
	\label{fig:3D}
\end{figure}

With such a 3D model, it is also possible to investigate the impact of certain inherently 3D structures on the performance of the vertical stability system. In particular, a \texttt{COMSOL} model of the VSC conductor and supporting structures was used to evaluate the induced current that can weave through the conductor jacket and mounting structure. The results of this study are presented in figure~\ref{fig:3D}, which shows the radial magnetic field ($B_\mathrm{r}$ -- measured at the magnetic axis) and $I_\mathrm{VSC}$ responses to a fixed VSC voltage waveform for SPARC configurations with and without supporting structures for the VSC conductors. It is immediately evident that the inclusion of these supporting structures comes at a heavy cost in VSC effectiveness: the peak current nearly doubles when the supports are included and the $B_\mathrm{r}$ response drops by almost a factor of two within this first $5\,$ms. Analysis of current patterns within the \texttt{COMSOL} model reveals that this degradation in performance stems primarily from the eddy current pattern created by inclusion of the supporting structures, where current travels directly from the large conductors in the vacuum vessel and supports into the jacket. With this insight, insulating layers were designed for the VSC supports that cut the conduction path from the jackets to the large plasma facing plates and thereby disrupt eddy currents in the system. Lower voltage specifications and shorter latencies can also contribute to reducing the eddy current inefficiencies in this system, yielding a final engineering design that meets the physics requirements outlined in section~\ref{sec:active}.

Similarly, a 3D model for a passive runaway electron mitigation coil (REMC) on SPARC \cite{tinguely_modeling_2021, izzo_runaway_2022} was included in characterizations of the effectiveness of the VSC systems. Again, it was determined that insulating the RMEC from secondary supports is critical for reducing eddy currents and maintaining maximum VSC effectiveness, as measured by the time response of $B_\mathrm{r}$ and thte required $I_\mathrm{VSC}$. 


\section{Summary and Outlook}
\label{sec:conc}

A series of 2D and 3D modeling, primarily with the \texttt{TokSys} code suite, is used to assess the prospects of vertical stability control on the SPARC tokamak and to investigate implications of this control for engineering decisions during design and construction of the machine. Because of its relatively thick conducting wall, which was designed to withstand disruption forces at high magnetic field and current, the passive stability of SPARC equilibria near the nominal operating point designed for $Q>1$ operation is already quite high. Most equilibria in this space have growth rates $\gamma<100\,$s$^{-1}$, with the elongation $\kappa_\mathrm{areal}$ playing the largest leveraging role in further reducing $\gamma$ to values in the range of $\gamma\sim30\,$s$^{-1}$. This optimistically suggests that vertical instability should not be a primarily limiting factor for SPARC operation. Further, the thick wall of SPARC creates an environment in which the addition of steel vertical stability plates in the near the divertors leads to only a modest ($25\%$) reduction on the equilibria growth rate while slightly reducing the effectiveness of the VSC system. 

Using a simple vertical control algorithm, the maximum controllable displacement is calculated to be in the range of $\sim6-7\%$ of the minor radius, which corresponds to safe expected operation as most seed perturbations should be on the order of $3\%$ of $a$, with larger perturbations up to $\Delta Z/a\sim5\%$ possible if sources are compounded or if very large ELMs (which should anyways be avoided) are produced. We note that, while the maximum controllable displacement is a good metric for evaluating the performance of the vertical control system, particularly when making comparisons between different systems, it does not capture the full story of plasma evolution. With close-fitting walls, as is targeted for SPARC and ARC, the metric that drives design choices is the ability to prevent the plasma from limiting following a vertical perturbation. 

This realization motivated a ``controllability” analysis, which utilized the same workflow to study the impact of the maximum VSC current, the maximum VSC voltage and the filter time constant on the vertical controllability of various equilibria families under different conditions. This process reveals a range of acceptable engineering parameters with different operation trade offs, allowing flexibility in the selection of machine components during SPARC design. An analysis of rigid displacements to families of equilibria around the nominal operating scenario is used to inform selection of the PFC contour shape, whereas higher-fidelity modeling with the \texttt{COMSOL} code reveals that the strategic inclusion of insulating layers with the VSC and REMC support structures are necessary to mitigate the impact of large eddy currents that would otherwise hinder the VSC response. 

As construction of the SPARC tokamak continues, so too will the development of more detailed models capturing the physics of vertical control. In particular, a nonlinear, time-dependent framework for control and simulation is being constructed that will be capable of capturing plasma evolution throughout entire modeled SPARC pulses. The backend of this framework uses \texttt{FGE}/\texttt{MEQ} to model the plasma and circuit dynamics \cite{langelaan_inter-code_2021}. When coupled with models of the internal plasma behavior provided by transport codes like \texttt{RAPTOR} \cite{felici_real-time-capable_2018}, this will allow for investigations of how other equilibrium effects will interact with vertical stability in a dynamic manner. Construction of this system will also enable explicit testing of vertical control algorithms, which can often compete with shape control demands, prior to SPARC operation. Finally, a comprehensive collection of off-normal event (ONE) models is currently under construction as part of the disruption assessment effort. Vertical stability (in particular due to large rapid displacements $Z_\mathrm{0}$) will be included in this collection as an initial test of the ONE handling architecture. Together, this extensive set of predict-first modeling should allow for a faster and more robust initiation of the SPARC tokamak.  


\section*{Acknowledgments}

This work was supported by Commonwealth Fusion Systems. 



\end{document}